\documentclass[letterpaper]{article} 
\usepackage{aaai2026}  
\usepackage{times}  
\usepackage{helvet}  
\usepackage{courier}  
\usepackage[hyphens]{url}  
\usepackage{graphicx} 
\usepackage{natbib}  
\usepackage{caption} 
\usepackage{algorithm}
\usepackage{algorithmic}
\usepackage{amsmath}
\usepackage{xcolor}
\usepackage[table]{xcolor}
\usepackage{booktabs}
\usepackage{multirow}
\usepackage{amssymb}
\usepackage{dsfont}
\usepackage{newfloat}
\usepackage{listings}
\DeclareCaptionStyle{ruled}{labelfont=normalfont,labelsep=colon,strut=off} 
\floatstyle{ruled}
\newfloat{listing}{tb}{lst}{}
\floatname{listing}{Listing}
\nocopyright

\title{Should You Use LLMs to Simulate Opinions?\\Quality Checks for Early-Stage Deliberation}

\author {
    Terrence Neumann\textsuperscript{\rm 1},
    Maria De-Arteaga\textsuperscript{\rm 2},
    Sina Fazelpour\textsuperscript{\rm 3}
}
\affiliations {
    \textsuperscript{\rm 1}McCombs School of Business, University of Texas at Austin\\
    \textsuperscript{\rm 2}Universitat Ramon Llull, ESADE\\
    \textsuperscript{\rm 3}Department of Philosophy and Computer Science, Northeastern University\\
}

\begin{document}

\maketitle

\begin{abstract}
The emergent capabilities of large language models (LLMs) have prompted interest in using them as surrogates for human subjects in opinion surveys. However, prior evaluations of LLM-based opinion simulation have relied heavily on costly, domain-specific survey data, and mixed empirical results leave their reliability in question. To enable cost-effective, early-stage evaluation, we introduce a quality control assessment designed to test the viability of LLM-simulated opinions on Likert-scale tasks without requiring large-scale human data for validation. This assessment comprises two key tests: \emph{logical consistency} and \emph{alignment with stakeholder expectations}, offering a low-cost, domain-adaptable validation tool. We apply our quality control assessment to an opinion simulation task relevant to AI-assisted content moderation and fact-checking workflows---a socially impactful use case---and evaluate nine LLMs using a baseline prompt engineering method (backstory prompting), as well as fine-tuning and in-context learning variants. None of the models or methods pass the full assessment, revealing several failure modes. We conclude with a discussion of the risk management implications and release \texttt{TopicMisinfo}, a benchmark dataset with paired human and LLM annotations simulated by various models and approaches, to support future research.
\end{abstract}

\begin{links}
    \link{Code \& Data}{https://tinyurl.com/qualitychecksAAAI}
    \link{Full Paper with Appendices}
    {https://arxiv.org/abs/2504.08954}
\end{links}

\section{Introduction}
Large language models (LLMs) have demonstrated significant appeal and versatility across a wide range of tasks. While the primary training objective of the base model is to predict the next most likely token, training on vast and diverse datasets, combined with increasingly complex system architectures, has resulted in the ``emergence'' of capabilities that were not explicitly anticipated during development \cite{wei2022emergent}. For instance, LLMs have been shown to learn new tasks through in-context learning \cite{brown2020fewshot}, perform coding tasks with remarkable accuracy \cite{zheng2024survey}, and have displayed human-like capabilities across different tasks \cite{kosinski2024evaluating, strachan2024testing}.

These emerging capabilities have led researchers to investigate more unconventional uses of LLMs, including automating opinion surveys by simulating the viewpoints of specific demographic or ideological groups \cite{qu2024performance}. The possibility that LLMs may successfully simulate demographic viewpoints stems from (i) evidence that LLMs can be prompted to replicate responses from human subject studies in psychology and other social sciences~\cite{aher2023simulate}, (ii) training data that likely encompass a broad spectrum of perspectives and opinions \cite{miranda2024diversity}, and (iii) prompting techniques that lead LLMs to produce seemingly coherent, opinionated responses \cite{ wright-etal-2024-llm}. Automating such surveys could have broad applications in marketing~\cite{sarstedt2024marketing}, content moderation~\cite{frohling2024personas}, policymaking, and public relations~\cite{sanders2023demonstrations}. LLMs may be especially valuable for pilot studies~\cite{sarstedt2024marketing, rothschild2024opportunities}, sampling hard-to-reach populations~\cite{jansen2023employing}, and labeling data that may be psychologically harmful for human annotators~\cite{wang2025large}. 

However, existing evidence is inconclusive on the extent to which LLM-based methods can accurately simulate human opinions. Some studies show that including a demographic ``backstory'' in a prompt \cite{Argyle2023simulate, bui2025mixture, jiang2024donald} (i.e., ``backstory'' prompting), fine‑tuning \cite{namikoshi2024using}, and in‑context learning \cite{karanjai2025synthesizing} are approaches that can successfully approximate group‑level opinions, while others document systematic failures of these same approaches---especially when simulating the opinions of minority or non‑Western groups \cite{santurkar2023aligning, sun2023aligningwhomlargelanguage, qu2024performance, mingmeng2024large, orlikowski2025beyond}. Additional work finds that backstory prompting approaches---the most prevalent approach to simulating opinions in prior research---flatten within-group variance of opinions \cite{wang2025large, Bisbee2024Synthetic, mingmeng2024large} and fails to generalize across topics \cite{sanders2023demonstrations, lee2024can}.  See \textbf{Appendix A} for a literature review table that summarizes prior works  methodologies, respective application domains, and findings.

The mixed results in the empirical literature suggest that the reliability of LLM-simulated opinions cannot be assumed across domains and instead requires rigorous validation for each new use case. Importantly, however, current validation methods rely on gathering high-quality human-labeled data---the cost of which can rival that of traditional surveys. For early‑stage assessments, this upfront expense is likely to outweigh uncertain benefits, significantly hindering research and development efforts. The existence of applications with broad consensus about the potential desirability of simulating opinions, such as when human survey subjects would be exposed to harmful content \cite{wang2025large} or trade secrets \cite{wang2024large}, motivates the need for validation techniques that lower the cost of early-stage exploration of LLM usage to simulate opinions.

\paragraph{Contribution 1: Designing a quality control assessment for early-stage deliberation of LLM opinion simulation.}
 
We introduce two diagnostic tests that probe desirable structural properties of simulated opinions without requiring the collection of large-sample human reference data:

\begin{enumerate}
    \item \textbf{Logical Consistency}---whether models reliably position ``average'' opinions as a convex combination of group-level opinions \emph{and/or} whether these average opinions are consistent with sampling from a stable reference population. 
    \item \textbf{Alignment with Stakeholder Expectations}---whether models position differences in group opinions in ways that are consistent with prior domain knowledge \emph{and/or} small sample survey data.
\end{enumerate}

The design of the proposed quality control assessment is grounded in the fact that some of the challenges involved in validating LLM-simulated opinions are similar to those involved in validating subjective human annotations. Researchers have developed various quality controls to assess the reliability of human data and annotator attentiveness \cite{artstein2008inter, sap-etal-2022-annotators, davani-etal-2022-dealing, lease2011quality}. Our proposed tests draw from this literature, adapting it to be suitable for assessing LLMs, with a focus on Likert-scale opinions. In the discussion section, we provide recommendations for integrating these tests into organizational risk management practices. 

The value of these tests compared to those proposed in previous work \cite{aher2023simulate, Argyle2023simulate, sun2023aligningwhomlargelanguage} is that they can be performed \emph{prior} to expensive large-scale human validation effort, fostering well-informed deliberation on the suitability of LLM opinion simulation in a particular context. 

\paragraph{Contribution 2: Crafting a domain-specific testbed.}

To illustrate the utility of our quality checks, we develop a domain-specific testbed in content moderation, focusing on the prioritization of potentially harmful misinformation for fact-checker review. Because prioritization is opinion-driven, resource-intensive, and exposes annotators to psychologically taxing content~\cite{liu2024exploring}, it provides a compelling, socially impactful context for evaluating simulated opinions.

An effective testbed for group-level opinion simulation should include topics with both expected agreement and disagreement amongst groups. Gender offers strong prior expectations based on a significant body of public opinion research~\cite{huddy2008gender, lizotte2020gender}. We focus on the prioritization criterion of identifying ``potential harm to specific groups,'' as previous research highlights this dimension of prioritization as one along which gender differences are particularly pronounced ~\cite{huddy2008gender, lizotte2020gender}.\footnote{While not the only criteria relevant for prioritization, research has noted that this is an important axis of consideration used by fact-checkers \cite{sehat2024harm, liu2024exploring}.}

We built \texttt{TopicMisinfo}, a 160‑claim dataset spanning topics that can be expected to elicit gender disagreement and consensus. The selection of these topics is anchored in findings from the American National Election Studies (ANES) \cite{ANES2021}, which help us capture a diversity in topics that are likely to provoke varying degrees of gendered disagreement. We also included ``Gold'' comprehension checks, which are claims that are harmless, and are either obviously true (e.g., ``A circle is round.'') or obviously false (e.g., ``The tallest tree on Earth touches Mars.''), with a very strong prior on gender-level consensus of opinion. We also collect a small sample ($\approx 1600$ annotations) of task-specific human opinion data. Dataset composition, example claims, and human annotation statistics are detailed in the \textbf{Appendices B-C}; the \texttt{TopicMisinfo} dataset is publicly available to support future research.

\paragraph{Contribution 3: Benchmarking leading models and approaches to opinion simulation using our quality checks.}

Applying the proposed quality checks, we evaluate simulated opinions from several commercially available LLMs: GPT-3.5-Turbo (checkpoints 06-13-23, 11-06-23, 01-25-24), GPT-4, GPT-4.1, GPT-5-mini, LLaMA 3, Titan-Text-Premier, and Mistral-Large---referred to throughout as GPT3.5a, GPT3.5b, GPT3.5c, GPT-4, GPT-4.1, GPT-5-mini, LLaMA3, Titan, and Mistral. We elicit group-level responses via backstory prompts~\cite{Argyle2023simulate, sun2023aligningwhomlargelanguage} that instruct models to adopt personas of men or women (see \textbf{Appendix D} for prompt templates). We also benchmark two other approaches to simulating opinions suggested in the literature: fine-tuning on small-sample human survey data~\cite{namikoshi2024using}; and in-context learning with human-labeled examples~\cite{karanjai2025synthesizing}. Importantly, our findings show that while fine-tuning and in-context learning generally improve alignment with stakeholder expectations, they do \textit{not} substantially outperform backstory prompting on logical consistency. These results indicate that all evaluated models and prompting approaches fail the proposed quality control assessment for the task considered, suggesting that current methods face limitations in their ability to produce stable, internally coherent simulations of human opinions.

\section{Quality Checks \& Results}
In this section, we introduce two quality checks designed for early-stage evaluation of LLM-simulated opinions. For each check, we detail the methodology and analyze how a range of LLMs perform across gender conditions using the ``backstory prompting'' approach \cite{Argyle2023simulate} on claims drawn from the \texttt{TopicMisinfo} dataset, focusing on a key opinion task relevant to fact-checking and content moderation workflows: \emph{assessing the potential harm of claims to specific groups.} The code and reproducible pipeline are publicly available.

\begin{figure*}[!ht]
  \centering
  \includegraphics[width=\textwidth]{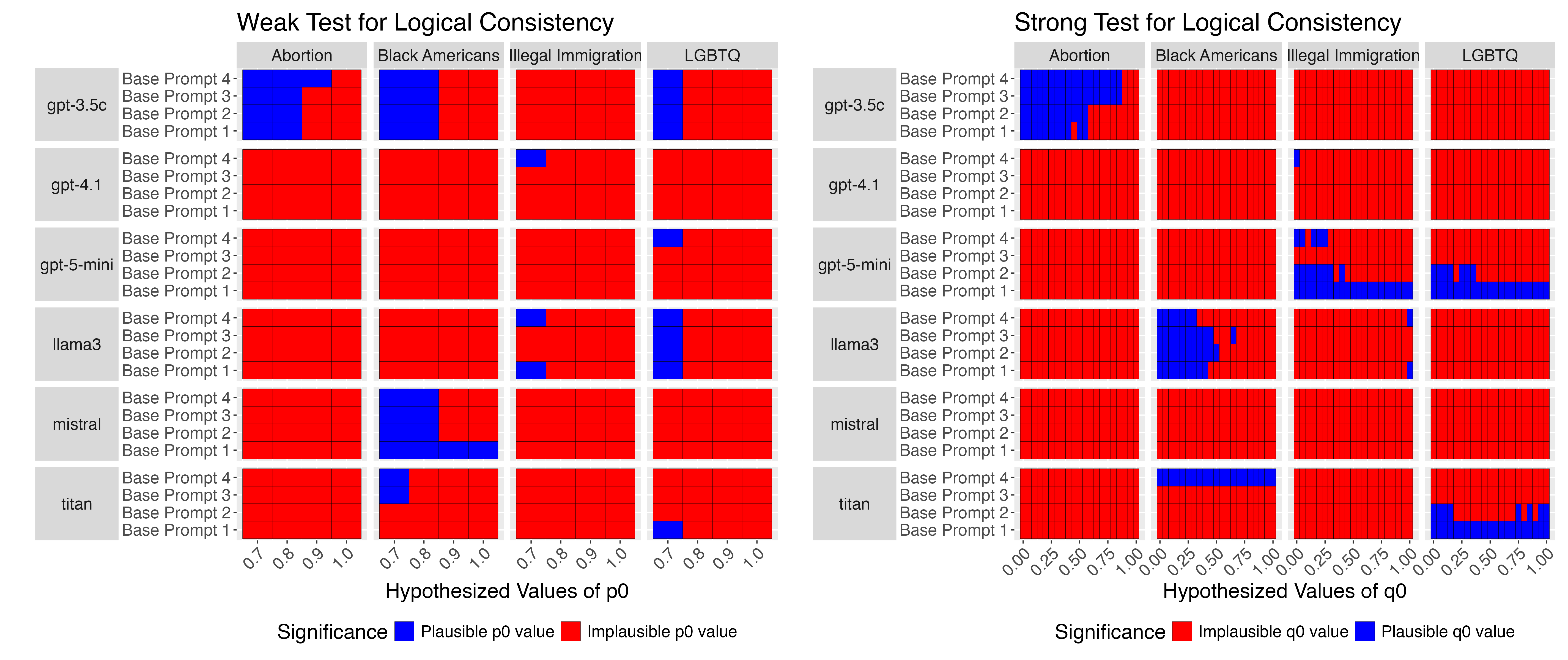}
  \caption{Results for weak and strong tests of logical consistency.}
  \label{fig:rq1}
\end{figure*}

\subsection{Quality Check 1: Logical Consistency}\label{subsec:exp1}

Testing for logical consistency is a foundational tool for evaluating the quality of data from human annotators \cite{aruguete2019serious} and from data aggregation methods such as majority voting \cite{davani-etal-2022-dealing}. In this work, we adapt this principle to assess the internal coherence of LLM-simulated opinions across demographic groups. 

When simulating opinions for multiple groups, a logically consistent model should produce an average opinion that is interpretable as a convex combination of the group-level opinions. This requirement reflects a basic expectation of distributional coherence: the ``average'' simulated opinion should be derivable from the underlying subgroups it is averaging over. For example, if in simulating the opinion of some population, $G$, consisting of two sub-groups, $\{g_1, g_2\}$, and the model simulates that members of $g_1$, on average, rate a claim's harmfulness as 2-out-of-6 and members of $g_2$, on average, as 4-out-of-6, then simulating an average opinion of 5-out-of-6 for $G$ as a whole is not just implausible, it is \emph{logically inconsistent}, as it violates a foundational algebraic constraint on how means combine.

Note that this is a rather weak constraint in the sense that it simply verifies that the average opinion for \textit{a given claim} falls between group-level opinions for that same claim. That is, the test allows the presumed proportion of the sub-populations, and so the weight given to their simulated average opinions, to vary \textit{across different claims} within a topic. A stronger test can, therefore, require that there exists a \emph{single, stable mixture weight}---e.g., a 50/50 blend of $g_1$'s and $g_2$'s opinions---that can consistently reconstruct the average opinion across all claims within a topic. A model that meets this condition supports the existence of a coherent reference population that the prompt reliably reflects. By contrast, failure to meet this criterion suggests that it is mathematically impossible for all simulated opinions across claims to reflect the views of a single population, casting doubt on how the simulated opinions can be interpreted and used.

These considerations can be summarized in terms of the following research questions:
\begin{itemize}
    \item \textbf{RQ-1a (Weak Test).} Does the model’s simulated average opinion fall within the convex hull of its group-conditioned opinions on at least a prespecified proportion \(p_0\) of claims in a given topic?
    
    \item \textbf{RQ-1b (Strong Test).} Is there a fixed mixture weight \(q_0 \in [0,1]\) that consistently reconstructs the model's average opinion across claims, indicating the presence of a stable underlying reference population?
\end{itemize}

\subsubsection{Methodology}

For each claim \(c_i\), the model is prompted \(n_i\) times under each condition (men, women, average) using a temperature setting of \(\tau = 0.5\), generating distributions of Likert-scale responses. A logically consistent average opinion should be expressible as a convex combination of the group-level opinions it seeks to summarize. In the context of gender (assuming only two: men and women)\footnote{While we assume a binary gender, \textbf{Appendices F and G} contain pseudo-code for generalizing this methodology for $\geq 2$ groups, while \textbf{Appendix H} includes the detail of a replication study including non-binary genders simulated with GPT3.5c.}, that would entail for each claim \(c_i\), the model's average predicted mean \(\mu_{\text{avg}}(c_i)\) should satisfy:
\[
\mu_{\text{avg}}(c_i) = \hat{q}_{c_i} \cdot \mu_{\text{men}}(c_i) + (1 - \hat{q}_{c_i}) \cdot \mu_{\text{women}}(c_i),
\]
for some \(\hat{q}_{c_i} \in [0, 1]\). If this condition holds, the average-conditioned response  can be interpreted as a plausible mixture of the two. If not, the average-conditioned response is logically inconsistent with the group-level outputs.

\subsubsection{Weak Test}

To assess logical consistency, we define the implied mixture weight \(\hat{q}_{c_i}^{(b)}\) for each bootstrap replicate \(b\) as:
\[
\mu_{\text{avg}}^{(b)}(c_i) = \hat{q}^{(b)}_{c_i} \cdot \mu_{\text{men}}^{(b)}(c_i) + (1 - \hat{q}^{(b)}_{c_i}) \cdot \mu_{\text{women}}^{(b)}(c_i)
\]
where the bootstrapped means \(\mu^{\text{group}}_{b}(c_i)\) are computed by resampling with replacement $n_i$ times from the original set of labels for each group. The mixture weight \(\hat{q}^{(b)}_{c_i}\) is then used to determine whether the average-conditioned mean lies within the convex hull of the group-specific means:
\[
I_i^{(b)} = 
\begin{cases}
1 & \text{if } \hat{q}^{(b)}_{c_i} \in [0, 1] \\
0 & \text{otherwise}
\end{cases}
\]

We aggregate the claim-level indicators within topic \(\omega\) to produce a topic-level success rate for each bootstrap sample:
\[
\hat{P}_\omega^{(b)} = \frac{1}{N_\omega} \sum_{i=1}^{N_\omega} I_i^{(b)}
\]

We then test the one-sided hypothesis:
\[
H_0\colon \hat{P}_\omega \geq p_0 
\quad \text{vs.} \quad 
H_1\colon \hat{P}_\omega < p_0
\]

The empirical \(p\)-value over \(B = 10^4\) replicates is computed as:
\[
\hat{p}(p_0) = \frac{1}{B} \sum_{b=1}^B \mathds{1}\left[\hat{P}^{(b)}_\omega \geq p_0\right]
\]

We reject \(H_0\) at level \(\alpha\) if \(\hat{p}(p_0) < \alpha\), concluding that the model fails to produce average-conditioned opinions that lie within the convex hull of group-level distributions at an acceptable rate. We test acceptable rate values \(p_0 \in [0.7, 0.8, 0.9, 1.0]\). See \textbf{Appendix F} for pseudocode that generalizes this test for $\geq 2$ groups.

\subsubsection{Strong Test}
To assess whether the average-conditioned responses could reflect a single, stable underlying population, we evaluate whether a fixed mixture weight \(q_0\) can consistently explain the simulated average-conditioned responses. For each claim \(c_i\), we compute the absolute deviation from a specified \(q_0 \in [0, 1]\):
\[
\hat{L}_\omega(q_0)
=
\frac{1}{N_\omega}
\sum_{i=1}^{N_\omega}
\left| \hat{q}_{c_i} - q_0 \right|
\]
Under the null hypothesis \(H_0\colon \hat{L}_\omega(q_0) = 0\), the model exhibits logical consistency at mixture weight \(q_0\).

We conduct a bootstrap hypothesis test with \(B = 10^4\) replicates. For each claim, we construct a synthetic mixture null hypothesis distribution:
\[
\mathcal{D}^{q_0}_{c_i} = q_0 \cdot \mathcal{D}^{\text{men}}_{c_i} + (1 - q_0) \cdot \mathcal{D}^{\text{women}}_{c_i}
\]
From this mixture, we resample synthetic average-conditioned responses, compute \(\hat{q}^{(b)}_{c_i}\), and aggregate across claims:
$\hat{L}^{(b)}_\omega(q_0)
=
\frac{1}{N_\omega}
\sum_{i=1}^{N_\omega}
\left| \hat{q}_{c_i}^{(b)} - q_0 \right|$. The \(p\)-value is $\Pr\left( \hat{L}^{(b)}_\omega(q_0) > \hat{L}_\omega(q_0) \right)$. This test is repeated for values of \(q_0\) in \([0, 1]\) at intervals of 0.05, applying a Bonferroni correction (\(\alpha^* = 0.0025\)) to account for multiple comparisons. See \textbf{Appendix G} for pseudocode that generalizes this test for $\geq 2$ groups.

\subsubsection{Results.}  Figure~\ref{fig:rq1} reports feasible consistency thresholds ($p_0$) for the weak test and valid $q_0$ intervals for the strong test, indicating alignment with a consistent reference population. We highlight results for four socially divisive topics where group-level disagreement is well-documented. Given the lack of a gold standard for prompting average-conditioned or population-level perspectives, we evaluate consistency across four alternative prompts. See \textbf{Appendix E} for details. 

\begin{itemize}
    \item \textbf{GPT-3.5c}: 
    For threshold $p_0=0.7$, passes the \textit{weak test} across three out of four topics with Base Prompt 4, indicating basic geometric plausibility. For the \emph{strong test}, it only passes the \emph{Abortion} topic, with feasible mixture weights \(q_0 \in [0, 0.55]\) or \([0, 0.85]\) depending on prompt. Fails to produce a consistent average-conditioned reference population for \emph{Black Americans}, \emph{Illegal Immigration}, and \emph{LGBTQ}.
    \item \textbf{GPT-4.1}: 
    One prompt passes the \textit{weak test} for \emph{Illegal Immigration}, but we see no other feasible thresholds across the prompts and topics tested. For the \emph{strong test}, we see the same prompt (Base Prompt 4) generate a plausible value for the topic of Illegal Immigration, but no other plausible values across topics and prompts.
    \item \textbf{GPT-5-mini}: The only reasoning model we tested does not perform markedly better. One feasible point for the \emph{weak test} for the \emph{LGBTQ} topic, but no others. We see better performance on the \emph{strong test} than other models, with feasible values for multiple prompts across two topics --- \emph{Illegal Immigration} and \emph{LGBTQ}.

    \item \textbf{LLaMA-3}: Two prompts pass the \emph{weak test} at \(p_0 = 0.7\) for \emph{Illegal Immigration} and all four prompts pass for \emph{LGBTQ}, but fails on other topics. For the \emph{strong test}, exhibits moderate success: feasible \(q_0 \in [0, 0.5]\) for \emph{Black Americans} across prompts, and up-samples opinions of men for \emph{Illegal Immigration} (\(q_0 \in [0.95, 1]\)). Fails on \emph{Abortion} and \emph{LGBTQ}.
    \item \textbf{Mistral}: Shows success on the \emph{weak test} only for \emph{Black Americans}, with feasible \(p_0\) values ranging from 0.7 to 1.0 depending on the prompt. Fails the \emph{strong test} across all topics—no consistent mixture weights were found.
    \item \textbf{Titan}: Two prompts pass the \emph{weak test} at \(p_0 = 0.7\) for \emph{Black Americans} and one prompt passes for \emph{LGBTQ} at $p_0 = 0.7$. On the \emph{strong test}, Prompt 4 yields success on \emph{Black Americans} with full-range plausibility (\(q_0 \in [0, 1]\)). Prompts 1 and 2 yield plausible \(q_0\ \in [0, 1]\) for \emph{LGBTQ}, but it fails on both \emph{Abortion} and \emph{Illegal Immigration}.
\end{itemize}

\paragraph{Take-away} Overall, most models fail to place average-conditioned opinions inside the convex hull of gendered responses at a reasonable rate, as evidenced by the results from the weak test. We see only \(\frac{13}{64} = 20.3\%\) and \(\frac{2}{64} = 3\%\) model$\times$topic$\times$prompt pairs passing a consistency threshold (\(p_0\)) of 0.8 and 0.9 respectively. Similarly, for the strong test, we see that models often do not often maintain a fixed reference population when sampling an average-conditioned opinion, as only \(\frac{13}{64} = 20.3\%\) of model$\times$topic$\times$prompt pairs reveal a fixed reference \(q_0\). 

\subsection{Quality Check 2: Alignment with Stakeholder Expectations and Small Sample Survey Data}\label{subsec:exp2}

Assessing alignment with common sense or widespread expectations is standard for evaluating human annotations, e.g., via attention checks \cite{lease2011quality}. The same underlying idea can be leveraged when assessing the quality of LLM-generated annotations. In our setting, for instance, stakeholders---e.g., end-users or developers---can have clear intuitions about cases where group-level differences should or should not arise. Checking for misalignment on these cases can provide a powerful lens for evaluating the quality of LLM-generated annotations. If neglected, such misalignment can bias downstream tasks, leading to skewed decisions or misrepresentation of group perspectives. In high-stakes settings like claim prioritization for fact-checking, such errors carry serious ethical consequences.

Rather than evaluating annotations individually, comparing simulated opinions across demographic groups can reveal critical early-stage information about the viability of opinion simulation. Group-level consensus on neutral topics and divergence on divisive ones offer early signals of realism and reliability. Quality Check 2 introduces a flexible method to test whether simulated group differences align with stakeholder expectations. For topics where domain experts anticipate clear consensus or pronounced disagreement across demographic groups (``clear-cut cases''), relying solely on stakeholder priors may be sufficient. In situations where expectations are less certain, small-scale human surveys serve as practical benchmarks. In our analysis, we leverage our own domain expertise as researchers in AI-assisted fact-checking to define stakeholder expectations and interpret test outcomes.

\paragraph{Research Questions.}
To what extent do simulated group differences in opinion conform to stakeholder expectations informed by (\textbf{RQ2a}) common-sense and domain-knowledge and (\textbf{RQ2b}) small-sample survey data?

\begin{table}
\centering
\begin{tabular}{lll}
\toprule
 & Topic: \textbf{Gold} & Topic: \textbf{Abortion} \\
 & \small (Prior: Insig. Differences) & \small (Prior: Sig. Differences) \\
\midrule
GPT-3.5a & \textcolor{red}{0.45***}  & \textcolor{blue}{3.42***} \\
GPT-3.5b & \textcolor{red}{0.36**}   & \textcolor{blue}{2.05***} \\
GPT-3.5c & \textcolor{red}{0.72***}  & \textcolor{blue}{2.37***} \\
GPT-4    & \textcolor{blue}{0.00}    & \textcolor{red}{0.07} \\
GPT-4.1  & \textcolor{red}{0.12**} & \textcolor{blue}{0.43**} \\
GPT-5-mini & \textcolor{red}{0.11***} & \textcolor{red}{-0.10} \\
llama3   & \textcolor{red}{-0.29*}    & \textcolor{blue}{0.13*} \\
mistral  & \textcolor{blue}{0.03}    & \textcolor{red}{0.53} \\
titan    & \textcolor{red}{0.33**}  & \textcolor{blue}{0.61*} \\
\bottomrule
\end{tabular}
\caption{Quality Check 2a Statistical Results — Alignment between model‑predicted gender differences and stakeholder priors. (\(*\)\,p$<$0.05; \(**\)\,p$<$0.01; \(***\)\,p$<$0.001).  \textcolor{blue}{Blue} values agree with the prior; \textcolor{red}{red} values contradict it. Results indicate that no model satisfies both priors.}
\label{tab:exp_3a}
\end{table}

\definecolor{poscolor}{RGB}{0,0,255}  
\definecolor{negcolor}{RGB}{255,140,0}  
\begin{table*}[h]
\captionsetup{skip=2pt}
\centering
\resizebox{\textwidth}{!}{%
    \begin{tabular}{lcccccccccccccccccc}
    \toprule
    Topic & \multicolumn{2}{c}{gpt-3.5a} & \multicolumn{2}{c}{gpt-3.5b} & \multicolumn{2}{c}{gpt-3.5c} & \multicolumn{2}{c}{gpt-4} & \multicolumn{2}{c}{gpt-4.1} & \multicolumn{2}{c}{gpt-5-mini} & \multicolumn{2}{c}{llama3} & \multicolumn{2}{c}{mistral-large} & \multicolumn{2}{c}{titan-text-premier} \\
 & $\bar{g}_{\omega}$ & p-val & $\bar{g}_{\omega}$ & p-val & $\bar{g}_{\omega}$ & p-val & $\bar{g}_{\omega}$ & p-val & $\bar{g}_{\omega}$ & p-val & $\bar{g}_{\omega}$ & p-val & $\bar{g}_{\omega}$ & p-val & $\bar{g}_{\omega}$ & p-val & $\bar{g}_{\omega}$ & p-val \\
\midrule
\multicolumn{19}{c}{\textbf{More Divisive Topics}} \\
\midrule
Abortion & \textcolor{poscolor}{1.65} & * & 0.28 & --- & 0.60 & --- & \textcolor{negcolor}{-1.70} & *** & \textcolor{negcolor}{-1.34} & *** & \textcolor{negcolor}{-1.89} & *** & \textcolor{negcolor}{-1.65} & *** & \textcolor{negcolor}{-1.24} & ** & \textcolor{negcolor}{-1.16} & * \\
Black Americans & \textcolor{poscolor}{1.72} & * & 0.35 & --- & 0.39 & --- & \textcolor{negcolor}{-1.28} & *** & \textcolor{negcolor}{-1.01} & *** & \textcolor{negcolor}{-0.92} & ** & \textcolor{negcolor}{-1.09} & *** & \textcolor{negcolor}{-1.92} & * & \textcolor{negcolor}{-0.97} & ** \\
Illegal Immigration & \textcolor{poscolor}{0.92} & ** & \textcolor{negcolor}{-1.06} & * & -0.61 & --- & \textcolor{negcolor}{-1.42} & ** & \textcolor{negcolor}{-1.05} & ** & \textcolor{negcolor}{-1.04} & ** & \textcolor{negcolor}{-0.80} & * & -0.43 & --- & \textcolor{negcolor}{-0.71} & * \\
LGBTQ & 0.66 & --- & \textcolor{negcolor}{-1.37} & * & -0.37 & --- & \textcolor{negcolor}{-0.92} & * & \textcolor{negcolor}{-1.06} & ** & \textcolor{negcolor}{-0.99} & * & \textcolor{negcolor}{-1.08} & ** & \textcolor{negcolor}{-0.99} & ** & \textcolor{negcolor}{-1.10} & ** \\
\midrule
\multicolumn{19}{c}{\textbf{Less Divisive Topics}} \\
\midrule
Entertainment & \textcolor{poscolor}{1.59} & * & \textcolor{poscolor}{1.05} & * & \textcolor{poscolor}{1.18} & * & \textcolor{poscolor}{0.35} & ** & \textcolor{poscolor}{0.39} & ** & \textcolor{poscolor}{0.47} & ** & \textcolor{poscolor}{0.41} & ** & \textcolor{poscolor}{0.58} & * & \textcolor{poscolor}{0.48} & * \\
Gold & 0.43 & --- & 0.34 & --- & \textcolor{poscolor}{0.71} & ** & -0.01 & --- & 0.10 & --- & 0.10 & --- & -0.30 & --- & 0.02 & --- & 0.32 & --- \\
HealthScience & \textcolor{poscolor}{1.86} & * & \textcolor{poscolor}{1.07} & * & 0.65 & --- & -0.19 & --- & \textcolor{poscolor}{0.58} & * & 0.32 & --- & 0.04 & --- & 0.79 & --- & 0.56 & --- \\
Sports & -0.01 & --- & -0.06 & --- & 0.21 & --- & -0.26 & --- & -0.26 & --- & -0.36 & --- & -0.26 & --- & -0.26 & --- & -0.16 & --- \\
USA & \textcolor{poscolor}{1.68} & *** & \textcolor{poscolor}{1.12} & *** & \textcolor{poscolor}{0.63} & ** & \textcolor{negcolor}{-0.61} & *** & -0.05 & --- & -0.15 & --- & -0.21 & --- & \textcolor{negcolor}{-0.56} & * & 0.16 & --- \\
WeatherClimate & \textcolor{poscolor}{1.06} & * & \textcolor{poscolor}{1.27} & * & \textcolor{poscolor}{0.97} & * & -0.39 & --- & 0.01 & --- & 0.06 & --- & -0.10 & --- & -0.68 & --- & 0.10 & --- \\

\bottomrule
\end{tabular}%
}

\footnotesize{Significance levels: \(*\) for p $<$ 0.05, \(**\) for p $<$ 0.01, \(***\) for p $<$ 0.001.\\
\textcolor{poscolor}{Blue}: positive significant values, \textcolor{negcolor}{Orange}: negative significant values.}
\caption{Quality Check 2b Statistical Results. GPT-3.5-turbo models tend to significantly exaggerate gender differences on topics humans found less divisive, while GPT-4, GPT-4.1, GPT-5-mini, Llama3, mistral-large, and Titan tended to significantly erode differences on topics humans found more divisive.}
\label{tab:exp_3b}
\end{table*}

\subsubsection{Methodology}

We evaluate whether LLM-generated gender opinion gaps align with stakeholder expectations or small-scale human surveys. For each claim \( c_i \) within topic \( \omega \), we define the model-estimated gender gap as:
\[
  \hat{D}(c_i) = \mu^{\text{woman}}_{\text{LLM}}(c_i) - \mu^{\text{man}}_{\text{LLM}}(c_i),
\]
and compare it to an expected gap \( E[D(c_i)] \), which is specified based either on stakeholder priors or small-scale surveys. We then compute the deviation $g_i = \hat{D}(c_i) - E[D(c_i)]$.

To create an efficient test statistic by reducing variance from imbalanced number of labels from simulated men and women, we weight each claim using the harmonic mean of the gender-group sizes:
\[
  w_i = \frac{n_i^{\text{woman}} \cdot n_i^{\text{man}}}{n_i^{\text{woman}} + n_i^{\text{man}}},
\]
a standard weighting approach in meta-analysis to manage group-size discrepancies and enhance estimator precision~\cite{hedges2014statistical}.

To test alignment at the topic level, we perform a weighted one-sample hypothesis test:
\[
  H_0: \bar{g}_\omega = 0 \quad \text{vs.} \quad H_1: \bar{g}_\omega \neq 0,
\]
where the weighted mean discrepancy for topic \( \omega \) is:
\[
  \bar{g}_\omega = \frac{1}{\sum_i w_i} \sum_{i=1}^{N_\omega} w_i \cdot g_i,
\]
and the appropriate weighted standard error \cite{nistweightsd} is:
\[
  \mathrm{SE}(\bar{g}_\omega) = \sqrt{\frac{\sum_i w_i (g_i - \bar{g}_\omega)^2}{(\sum_i w_i - 1)\sum_i w_i}}.
\]
This yields the topic-level \emph{t}-statistic $t_\omega = \frac{\bar{g}_\omega}{\mathrm{SE}(\bar{g}_\omega)}$, with degrees of freedom \( \nu = \sum_i w_i - 1 \).

Each model response in our dataset corresponds to a 6-point Likert-scale judgment, formally an ordinal variable. Nonetheless, parametric methods such as the \emph{t}-test can reliably analyze Likert-type data when averaged over moderate-to-large samples~\cite{norman2010likert, deWinter2010likert}. In our experiments, each model--gender--claim cell aggregates at least 10 LLM completions, and statistics are computed at the level of \emph{topic means}, averaged over individual claims, ensuring approximate normality via the Central Limit Theorem.

\paragraph{RQ-2a: Testing Against Stakeholder Priors.}

We select topics from \texttt{TopicMisinfo} known through domain expertise to be either clearly non-divisive (e.g., \textbf{Gold} attention checks) or clearly divisive (e.g., claims related to \textbf{Abortion}). For non-divisive topics, stakeholder intuition strongly supports minimal or no demographic differences; thus, we explicitly set \( E[D(c_i)] = 0 \). Similarly, for divisive topics, setting \( E[D(c_i)] = 0 \) serves as a falsification test: substantial deviations from zero confirm that the model appropriately simulates expected group divergences. Thus, failing to reject \( H_0 \) for non-divisive topics aligns with stakeholder priors, confirming minimal simulated group differences. Conversely, rejecting \( H_0 \) for divisive topics validates that simulated opinions mirror expected demographic divergences.

\paragraph{RQ-2b: Testing against human annotations.}
When human survey data are available, we define \( E[D(c_i)] = \mu^{\text{woman}}_{\text{Human}}(c_i) - \mu^{\text{man}}_{\text{Human}}(c_i) \). We then calculate $g_i = \hat{D}(c_i) - E[D(c_i)]$. Although the gender gaps are estimated from distinct sources (LLM vs. human samples), the test is structured as a paired one-sample analysis over the differences \( g_i \), with standard error and degrees of freedom computed identically. Rejection indicates meaningful divergence between LLM-simulated and human-annotated gender differences. 

\paragraph{Results (see Tables~\ref{tab:exp_3a}–\ref{tab:exp_3b}).}
\begin{itemize}
    \item \textbf{Exaggeration on consensus topics.} In Quality Check 2a, all models except GPT-4 and Mistral report spurious gender gaps on the \textbf{Gold} ``circle-is-round'' topic ($p < \alpha$), indicating oversensitivity to gender distinctions. In Quality Check 2b, the GPT-3.5 family of models exaggerate disagreement relative to human survey data on less polarizing topics like \textbf{entertainment}, \textbf{health \& science}, \textbf{U.S. politics}, and \textbf{climate}.
    \item \textbf{Erosion on divisive topics.} Other models---including GPT-4, GPT-4.1, GPT-5-mini, LLaMa3, Titan, and Mistral---tend to erode known gender differences, even on topics with well-established disagreement. Quality Check 2a shows that GPT-4, GPT-5-mini, and Mistral fail to reflect divergent opinions on \textbf{abortion}, and Quality Check 2b reveals that these models, in addition to GPT-4.1, Titan and LLaMa3, also underrepresent human gender-level disagreement on topics such as \textbf{Black Americans}, \textbf{LGBTQ} issues, and \textbf{Illegal Immigration}, yielding significant negative $\bar{g}_\omega$ scores ($p < \alpha$).
\end{itemize}


\paragraph{Take-away.}  
Current LLMs often misalign with stakeholder priors—some exaggerate less divisive claims, others erode salient divides. Further, there are clear differences between the models tested: the GPT-3.5-turbo variants tend to be guilty of exaggerating gender-level differences, while GPT-4, Llama 3, Mistral, and Titan tend to erode differences across topics, both when conditioned on prior expectations and on small sample survey data.

\section{Benchmarking Experiment}
In the previous section, we tested our quality checks using the backstory prompting approach across several different models. To assess alternative strategies, we benchmark this approach against two others prominent in the literature: fine-tuning \cite{namikoshi2024using} and in-context learning \cite{karanjai2025synthesizing}. Specifically, we use only the GPT-3.5c checkpoint to control for variation across model architectures and focus on comparing simulation strategies. We test the following strategies:

\begin{enumerate}
\item \textbf{Prompt Engineering}: Two variations of backstory prompts (``Cond. Prompt 1'' and ``Cond. Prompt 2''), which differ in framing and specificity of demographic identity, direct the model to simulate opinions from specific demographic perspectives. See \textbf{Appendix 4} for exact prompt specifications.
\item \textbf{Fine-Tuning}: We fine-tune GPT3.5c on the average \textbf{human} ratings for each condition (men, women, overall average) using 40 claims from \texttt{TopicMisinfo}, yielding 120 training examples. Fine-tuning is performed over three epochs. See \textbf{Appendix 6a} for more detail.
\item \textbf{In-Context Learning (ICL)}: We test two strategies: (a) \textbf{Random Sampling}, where prompts include random labeled examples for each condition (men, women, overall average), and (b) \textbf{Nearest Neighbor}, where prompts include the most semantically similar claims and their associated labels based on cosine similarity. See \textbf{Appendix 6b} for more detail on methodology and examples.
\end{enumerate}

\paragraph{Experimental Setup}
We partition the \texttt{TopicMisinfo} dataset into 120 held-out claims for evaluation and 40 claims for training or in-context injection. We reapply our baseline prompt engineering methods to the held-out set, fine-tune on human-labeled examples, and construct in-context prompts using the 40 available claims. Each method is evaluated using a summary metric of our quality checks. For QC1a, we report the proportion of $p_0$ thresholds passed across four topics. For all other checks (QC1b, QC2a, QC2b), we report the percentage of topics for which the method successfully satisfies the quality criterion.

\paragraph{Results}
Table~\ref{tab:benchmark_results} summarizes outcomes. Relative to the baseline approach of backstory prompting (Cond. Prompt 1 and Cond. Prompt 2), we find that: \textbf{Fine-tuning} achieves perfect alignment with priors and human data (QC2a/b = 100\%) but degrades logical coherence (QC1a/b); \textbf{Random-sample ICL} matches fine-tuning on QC2b and retains moderate logical consistency; \textbf{Nearest-neighbor ICL} delivers the best score on the weak test for logical consistency (QC1a = 88\%) yet dramatically fails the strong test (QC1b = 0\%). Notably, all methods struggle on QC1b, underscoring the open challenge of producing a stable ``average'' reference population.

\begin{table}[t]
\centering
\begin{tabular}{lcccc}
\toprule
Method & QC1a & QC1b & QC2a & QC2b \\
\midrule
Cond. Prompt 1          & 44\%          & 25\%           & 50\%             & 70\%  \\
Cond. Prompt 2          & 44\%          & \textbf{50\%}  & 0\%             & 70\%  \\
Fine-tuning             & 19\%          & 25\%           & \textbf{100\%}   & \textbf{100\%} \\
Random Sample (ICL)     & 56\%          & \textbf{50\%}  & 50\%             & \textbf{100\%}    \\
Nearest Neighbor (ICL)  & \textbf{88\%} & 0\%            & 50\%             & 90\%    \\
\bottomrule
\end{tabular}
\caption{Performance on quality checks across methods. Fine-tuning improves alignment, but not logical consistency compared to baseline.}
\label{tab:benchmark_results}
\end{table}

\section{Discussion}
LLMs offer a promising avenue for simulating demographic opinions efficiently without immediate reliance on costly human-generated data. However, ensuring these outputs truly capture nuanced human perspectives remains challenging, particularly because LLMs often produce superficially coherent but logically flawed responses. To address this, we introduce an early-stage assessment targeting logical consistency and alignment with stakeholder expectations. These tests can help organizations quickly evaluate whether simulated opinions from LLMs justify deeper investment and development in their particular application domain.

Applying our quality checks, we uncovered significant logical inconsistencies: specifically, most models (80\%) produced ``average'' opinions more extreme than demographic-specific predictions, violating basic statistical logic. Unlike logical inconsistencies of LLMs exposed in previous research---for example, mutually incompatible factual beliefs \cite{kassner2021beliefbank}, self‑inconsistent chain‑of‑thought traces \cite{wang2022selfconsistency}, violations of basic propositional logic \cite{ghosh2024logic}, or within‑passage contradictions \cite{mundler2024selfcontradictions}---these errors indicate fundamental logical flaws 
at the distributional level. 
Practitioners cannot simply acknowledge and proceed despite these errors; they must reconsider or even abandon flawed simulation strategies entirely.

Further, our checks reveal systematic discrepancies with stakeholder expectations. Older models (like GPT-3.5-turbo) tend to exaggerate opinion differences even on obviously non-divisive claims, risking harmful stereotyping or inefficient resource allocation. Conversely, newer models (GPT-4, GPT-4.1, GPT-5-mini, LLaMa3, Mistral, Titan) often obscure genuine demographic divides on controversial topics, potentially masking critical areas affecting particular groups. Importantly, we find that the magnitude of gender-level differences observed in QC2a correlates with the direction of the errors observed in QC2b, suggesting that testing clear-cut cases aligned with stakeholder priors---even in the absence of extensive human data---provides a valuable preliminary quality check of LLM reliability.

Finally, in a benchmarking experiment, we found that fine-tuning and in-context learning generally improve alignment with stakeholder expectations, but they do not substantially outperform backstory prompting in terms of generating a stable reference population (QC1b). These results suggest that current methods remain limited in their ability to produce stable, internally coherent simulations, highlighting a promising avenue for future work.

\paragraph{Limitations}

Our study provides a reusable framework for early‐stage evaluation of LLM‐simulated opinions, but several factors define the boundaries of its current scope and suggest directions for future work. First, in our empirical results we benchmark exclusively instruction–tuned language models, as prior literature suggests that such models are likely to perform better than others on instruction–style prompts. The proposed tests are agnostic to model type, but we leave evaluation of base models to future work. Second, our experiments use only two binary gender personas and an unprimed ``average'' persona as proof of concept. We constrain ourselves to binary gender due to limitations of the human data, and in Appendix H we show how QC1 can be extended to non-binary gender. Crucially, in many application domains the relevant demographic or psychographic spectrum is much richer. Identifying all important subgroups and capturing their diversity remains an open challenge; omitting a salient group can invalidate the average–persona check and obscure meaningful variation. Finally, the human annotator pool for our TopicMisinfo dataset is predominantly male (\(\approx 69\,\%\) men, \(31\,\%\) women).  Although QC2b uses only subgroup–specific means and is therefore unaffected by the imbalance, this gender distribution could influence population‐level statistics if the dataset is reused for other purposes.

\paragraph{Risk Management: Logical Consistency} From a risk management perspective, the choice between weak and strong logical consistency tests should align directly with the intended use of LLM-simulated outputs. In preliminary evaluations or exploratory research, the weak test---which verifies that average responses for each claim lie within the range defined by subgroup opinions---might suffice as a basic sanity check. However, in high-stakes resource-allocation scenarios such as prioritizing claims for fact-checking or deciding which content requires urgent human moderation, the strong test becomes crucial. For example, if a fact-checking organization uses a simulated ``average'' opinion as a baseline to determine whether specific demographic groups are disproportionately affected by misinformation, the average stance must represent a consistent, interpretable reference population. Inconsistencies in this average opinion could obscure whether certain groups genuinely require attention, leading to misallocation of limited fact-checking resources. Thus, adopting the strong test in such settings ensures stable benchmarks, supports defensible allocation decisions, and enhances overall reliability in high-consequence decision-making processes.

\paragraph{Risk Management: Alignment with Stakeholder Expectations} Organizations should prioritize testing whether model outputs align with clear common-sense or domain-specific expectations, paying close attention to discrepancies with their justified priors. For instance, an error of exaggeration of differences between groups on a non-divisive issue is likely a more serious error than an error of exaggeration on an issue known to be divisive.  Moreover, while our current metrics focus on \emph{mean differences} in opinion distributions, some stakeholders may prioritize other distributional properties---such as variance or skew of opinions \cite{bui2025mixture, wang2025large}. Our framework is flexible: the test statistics we propose can be adapted to target other distributional moments.

\paragraph{Conclusion} By applying quality control checks that surface logical consistency and alignment with stakeholder expectations, practitioners can determine whether an approach to opinion simulation clears a basic threshold for reliability before devoting significant resources to collecting large-scale human reference data. An approach that meets these conditions may still require subsequent testing and tuning, but it at least demonstrates fundamental viability. If it fails on multiple fronts, stakeholders should carefully consider whether an LLM-driven approach is worthwhile or whether alternative data-centric or human-led solutions would be more consistent and trustworthy.

\section*{Acknowledgments.} This work was supported in part by Good Systems, a research grand
challenge at the University of Texas at Austin. S.F. acknowledges support from the Schmidt Sciences AI2050 Early Career Fellowship. 

\clearpage

\onecolumn

\definecolor{lightgreen}{HTML}{D9EAD3}
\definecolor{lightyellow}{HTML}{FFF2CC}
\definecolor{lightred}{HTML}{F4CCCC}

\section{Appendix A: Literature Review Table}

\begin{table}[!ht]
\footnotesize
\centering
\caption*{Appendix Table 1: Overview of studies using LLMs to simulate human opinions, grouped by task, approach, dataset used for evaluation, LLM used, and key findings. Color coded by author-specified outcome of quality of simulated opinions (e.g., good, bad, neutral).}
\label{tab:llm_opinion_table}
\begin{tabular}{p{3.2cm} p{2.5cm} p{3cm} p{1.5cm} p{1.2cm} p{3cm}}
\toprule
\textbf{Title} & \textbf{Task} & \textbf{Approach} & \textbf{Dataset} & \textbf{LLMs} & \textbf{Key Findings} \\
\midrule

\rowcolor{lightgreen}
\textit{Out of One, Many} \cite{Argyle2023simulate} &
Political Attitudes &
\textbf{Prompt Engineering}: Backstories &
ANES &
GPT-3 &
\textbf{Positive}. Useful for simulating opinion distributions. \\
\hline
\rowcolor{lightgreen}
\textit{Donald Trumps} \cite{jiang2024donald} &
Voting &
\textbf{Prompt Engineering}: Backstories + structural &
ANES, WVS &
ChatGPT &
\textbf{Positive}. Backstory prompts produce outputs that match voting patterns. \\
\hline
\rowcolor{lightgreen}
\textit{Synthesizing Public Opinion} \cite{karanjai2025synthesizing} &
Political Attitudes &
\textbf{In-Context Learning}: Knowledge Injection &
CES &
Gemma2, LLaMa3 &
\textbf{Positive}. Incorporating knowledge in prompts improves over backstory. \\
\hline
\rowcolor{lightgreen}
\textit{Mixture of Personas} \cite{bui2025mixture} &
Sentiment &
\textbf{Prompt Engineering}: Probabilistic Mixture of Personas &
IMDB, Yelp, AGNews &
LLaMa3 &
\textbf{Positive}. Better alignment and within-group diversity than backstory prompting. \\
\hline
\rowcolor{lightgreen}
\textit{Modeling Targeted Populations} \cite{namikoshi2024using} &
EV attitudes &
\textbf{Fine-tuning}: ID \& Demographics &
BEV Survey &
LLaMa2 &
\textbf{Positive}. Fine-tuning improved KL-divergence and RMSE with subgroups. \\
\hline
\rowcolor{lightyellow}
\textit{Aligning with Whom?} \cite{sun2023aligningwhomlargelanguage} &
Politeness &
\textbf{Prompt Engineering}: Backstories &
POPQUORN &
GPT-3.5, GPT-4 &
\textbf{Mixed}. Aligns more with White participants than Black/Asian ones. \\
\hline
\rowcolor{lightyellow}
\textit{Whose Opinions?} \cite{santurkar2023aligning} &
Political Attitudes &
\textbf{Prompt Engineering}: Backstories &
OpinionQA &
GPT-3.5 &
\textbf{Mixed}. Matches younger/liberal demos; some steering helps. \\
\hline
\rowcolor{lightyellow}
\textit{Climate Opinion} \cite{lee2024can} &
Environmental attitudes &
\textbf{Prompt Engineering}: Backstories + covariates &
Novel. Nat'l sample &
GPT-3.5, GPT-4 &
\textbf{Mixed}. Demographics alone insufficient; psych vars improve results. \\
\hline
\rowcolor{lightyellow}
\textit{Synthetic Survey Data} \cite{Bisbee2024Synthetic} &
Feeling thermometers &
\textbf{Prompt Engineering}: Backstories &
ANES &
GPT-3.5-Turbo &
\textbf{Mixed}. Similar means, reduced variance; prompt/time sensitive. \\
\hline
\rowcolor{lightyellow}
\textit{Beyond Demographics} \cite{orlikowski2025beyond} &
Content moderation &
\textbf{Fine-tuning}: IDs \& Demographics &
MINT, POPQUORN, DICES &
LLaMa 3 &
\textbf{Mixed}. Models learn individual opinions, but fail at demographic-level. \\
\hline

\rowcolor{lightred}
\textit{Chameleons?} \cite{mingmeng2024large} &
Political Attitudes &
\textbf{Prompt Engineering}: Backstories &
ESS &
GPT-3.5, Llama 2/3, Mistral, Deepseek &
\textbf{Negative}. Highly sensitive to prompt; misrepresents group-level opinions. \\
\hline
\rowcolor{lightred}
\textit{LLMs Flatten Identities} \cite{wang2025large} &
Political Attitudes &
\textbf{Prompt Engineering}: Backstories &
Novel (n=3200) &
GPT-4, Llama 2, WizardLM &
\textbf{Negative}. Misrepresents minorities; flattens within-group variation. \\
\hline
\rowcolor{lightred}
\textit{Cross-Cultural Simulation} \cite{qu2024performance} &
Political Attitudes &
\textbf{Prompt Engineering}: Backstories &
WVS &
GPT-3.5-Turbo &
\textbf{Negative}. Performs better on U.S. samples; shows liberal bias. \\

\bottomrule
\end{tabular}
\end{table}

\clearpage

\section{Appendix B: Data Collection Procedure - \emph{TopicMisinfo} Dataset}

Using Amazon SageMaker Ground Truth, we hired American workers and compensated them at a rate of \$15-20 per hour to annotate various claims. We asked them to annotate each claim according to different important axes of prioritization. 
Specifically, we asked for annotations of a claim's along four dimensions: (1) ``Prioritization'': need for prioritization in fact-checking (based on their own subjective criteria); (2) ``General Public'': interest to the general public; (3) ``Group Harm'': potential to harm certain demographic groups; and (4) ``Perceived Falsehood'': perceived degree of veracity. Table \ref{tab:check} shows the questions that annotators were asked. If workers identified a claim as harmful to a specific group(s), we asked them to identify which group(s) may be harmed. After workers completed the task, we asked them to fill out a demographic survey. 

Annotated claims were distributed across nine salient topics pertinent to diverse American groups, namely: Health and Science, USA (politics), Illegal immigration, Abortion, LGBTQ issues, Weather and Climate, African Americans, Entertainment, and Sports. Our selection of topics was driven by their potential to elicit varied prioritizations and to mirror the multifaceted nature of society, where numerous significant information topics vie for attention. To source these claims, we employed Google’s Fact-Check API, filtering by the specified topics. This API consolidates recent fact-checks from reputable sources, including Snopes, PolitFact, and FullFact, ensuring that all claims in our dataset were from the preceding year. Importantly, each topic encompassed both accurate and inaccurate information, enabling us to gauge sensitivity to falsehood within specific topics and contexts. In total, we gathered 1998 annotations for each of the four definitions of check-worthiness spanning 160 distinct claims. 

To ensure the quality and reliability of our data, we introduced ``gold'' data as an attention check. These gold questions were either manifestly false (e.g.,``The largest tree creates a land bridge between the earth and Mars.'') or patently true (e.g., ``A circle is round''). We retained only the data from workers who demonstrated over 80\% accuracy in discerning the veracity of a claim when presented with at least two gold questions. Concurrently, we eliminated a small number of duplicate responses in which worker evaluated the same claim multiple times. This filtering yielded 1592 annotations from 27 distinct workers. 

\begin{table}[h]
\centering
\begin{tabular}{|l|p{10cm}|}
\hline
\textbf{Definition} & \textbf{Description} \\
\hline
``Prioritization'' & Due to limited resources, fact-checking organizations must \emph{prioritize} which claims to check. In your opinion, should this claim be prioritized for fact-checking? (1 - 6 scale)\\
\hline
``General Public'' & To what extent will the claim be of interest to the general public? (1 - 6 scale)\\
\hline
``Group Harm'' &  How likely is this claim to disproportionately harm certain demographic groups more than others (e.g. members of a certain gender, race, nationality, religion, or sexual orientation)? (1 - 6 scale)\\
\hline
``Perceived Falsehood'' & Does the claim appear to be completely true, completely false, or contain both true and false information? (1 - 6 scale, higher meaning more false)\\
\hline
\end{tabular}
\caption*{Appendix Table 2: Different Axes of Check-Worthiness Assessed.}
\label{tab:check}
\end{table}

\clearpage

\section{Appendix C: Example Claims from Dataset}

The following table demonstrates claims across topics in the TopicMisinfo Dataset.

\begin{table*}[!ht]
\centering
\begin{tabular}{|p{0.5\textwidth}|l|}
\toprule
\textbf{Claim} & \textbf{Topic} \\
\midrule
Abortion can cause permanent psychological and mental health damage, including post-abortion syndrome & Abortion \\ \hline
Food crops can be engineered right now based on existing technology to cause infertility in Black People alone & Black Americans \\ \hline
The character of James Bond has been portrayed by seven different actors over the course of the film franchise's 60-year history. & Entertainment \\ \hline
All diseases can be spread through radio waves. & GoldFalse \\ \hline
Paris is located in France. & GoldTrue \\ \hline
Ingesting Borax has health benefits & HealthScience \\ \hline
A Texas farmer was arrested for using illegal immigrants for free labor by pretending to be Joe Biden. & Illegal Immigration \\ \hline
In June 2022, deputies from the New Hanover Sheriff's Office "escorted" far-right Proud Boys protesters towards an LGBTQ Pride event for children at a library in Wilmington, North Carolina. & LGBTQ \\ \hline
The longest recorded field goal was made from a distance of 64 yards. & Sports \\ \hline
Biden asked China for permission to shoot down a spy balloon & USA \\ \hline
The state of Hawaii has the highest number of lightning strikes per square mile of any state in the United States. & WeatherClimate \\ 
\bottomrule
\end{tabular}
\caption*{Appendix Table 3: Example Claims and Their Associated Topics}
\end{table*}

\clearpage

\section{Appendix D: Descriptive Analysis of Dataset}

In this section, we provide several tables that demonstrate key descriptive statistics about the \texttt{TopicMisinfo} dataset.

\begin{table}[!ht]
\centering
\footnotesize
\begin{tabular}{|p{0.2\textwidth}|c|c|c|c|}
\hline
\textbf{Topic} & \textbf{\# FALSE Claims} & \textbf{\# TRUE Claims} & \textbf{\# Men Annotations} & \textbf{\# Women Annotations} \\
\hline
Abortion & 13 & 6 & 119 & 61 \\ \hline
Black Americans & 6 & 6 & 78 & 37 \\ \hline
Entertainment & 5 & 4 & 63 & 23 \\ \hline
Gold & 11 & 15 & 199 & 141 \\ \hline
Health and Science & 9 & 1 & 71 & 30 \\ \hline
Illegal Immigration & 11 & 3 & 88 & 37 \\ \hline
LGBTQ & 10 & 4 & 93 & 45 \\ \hline
Sports & 3 & 2 & 27 & 17 \\ \hline
USA & 38 & 5 & 280 & 120 \\ \hline
Weather and Climate & 3 & 5 & 42 & 21 \\ \hline
\hline
\textbf{Total} & \textbf{109} & \textbf{51} & \textbf{1060} & \textbf{532} \\
\hline
\end{tabular}
\caption*{Appendix Table 4: Distribution of TRUE and FALSE claims by topic, with number of annotations gathered by gender.}
\label{tab:distributions}
\end{table}

\begin{table}[!ht]
\centering
\begin{tabular}{|l|l|}
\hline
\textbf{Demographic} & \textbf{Distribution} \\
\hline
Gender & 69\% Male / 31\% Female \\
\hline
Race & 66\% White / 25\% Asian / 2\% Black / 6\% NA \\
\hline
Sexual Orientation & 86\% Straight / 14\% Gay and Bisexual \\
\hline
\end{tabular}
\caption*{Appendix Table 5: Demographic analysis of workers that created annotations for \texttt{TopicMisinfo} dataset (\% of claims labeled by group).}
\label{tab:demographics}
\end{table}

\begin{table}[!ht]
\centering
\label{tab:topic_gender_stats}
\begin{tabular}{|lrrrr|}
\toprule
\multirow{2}{*}{\textbf{Topic}} & \multicolumn{2}{c}{\textbf{Mean}} & \multicolumn{2}{c}{\textbf{Std. Dev.}} \\
\cmidrule(r){2-3} \cmidrule(r){4-5}
 & Female & Male & Female & Male \\
\midrule
Abortion            &  4.93 &  3.18 &  1.36 &  1.82 \\
Black Americans     &  4.86 &  3.73 &  1.58 &  1.82 \\
Entertainment       &  1.04 &  1.36 &  0.21 &  1.00 \\
GoldFalse           &  1.11 &  1.07 &  0.36 &  0.40 \\
GoldTrue            &  1.05 &  1.04 &  0.34 &  0.31 \\
HealthScience       &  1.30 &  1.25 &  0.75 &  0.86 \\
Illegal Immigration &  4.14 &  3.01 &  1.53 &  1.75 \\
LGBTQ               &  4.96 &  3.96 &  1.22 &  1.88 \\
Sports              &  1.41 &  1.07 &  1.23 &  0.27 \\
USA                 &  1.98 &  1.53 &  1.60 &  1.21 \\
WeatherClimate      &  1.33 &  1.07 &  1.06 &  0.34 \\
\bottomrule
\end{tabular}
\caption*{Appendix Table 6: Human Summary Statistics from Likert-Scale Ratings of Claims by Topic and Gender}
\end{table}

\clearpage

\section{Appendix E: Prompts for Experiments}

The following table demonstrates the prompts that we used across the various experiments. After experimentation with temperature parameter models revealed insignificant within-model change in rating or variability, we decided to set $\text{temperature } (\tau) = 0.5$ for all models across all prompts.

\begin{table*}[!ht]
\centering
\begin{tabular}{|p{4cm}|p{12cm}|}
\hline
\multicolumn{2}{|c|}{\textbf{Gender-Conditioned Prompts}} \\
\hline
\textbf{Prompt} & \textbf{Format} \\
\hline
\textbf{Conditional Prompt 1} & 
Claim: \{\textbf{claim}\} \newline
Answer the following question as if you were a \{\textbf{gender}\}: \newline
How likely is this claim to disproportionately harm certain demographic groups \newline
(e.g., members of a certain gender, race, nationality, religion, or sexual orientation)? \newline
(1-6 scale) \newline
As a \{\textbf{gender}\}, on a scale from 1-not at all likely to 6-extremely likely, I would rate it as a: \\
\hline
\textbf{Conditional Prompt 2} & 
Answer the following question as if you were a \{\textbf{gender}\}: \newline
How likely is this claim to disproportionately harm certain demographic groups \newline
(e.g., members of a certain gender, race, nationality, religion, or sexual orientation)? \newline
(1-6 scale) \newline
Options: 1-not at all likely, 2, 3, 4, 5, 6-extremely likely \newline
Claim: \{\textbf{claim}\} \newline
Response (Provide the answer without explaining your reasoning): \\
\hline
\multicolumn{2}{|c|}{\textbf{Statistically Neutral (Base) Prompts}} \\
\hline
\textbf{Base Prompt 1} & 
Claim: \{\textbf{claim}\} \newline
How likely is this claim to disproportionately harm certain demographic groups \newline
(e.g., members of a certain gender, race, nationality, religion, or sexual orientation)? \newline
(1-6 scale) \newline
On a scale from 1-not at all likely to 6-extremely likely, I would rate it as a: \\
\hline
\textbf{Base Prompt 2} & 
Claim: \{\textbf{claim}\} \newline
Answer the following question from \textbf{a neutral perspective}. \newline
How likely is this claim to disproportionately harm certain demographic groups \newline
(e.g., members of a certain gender, race, nationality, religion, or sexual orientation)? \newline
(1-6 scale) \newline
On a scale from 1-not at all likely to 6-extremely likely, I would rate it as a: \\
\hline
\textbf{Base Prompt 3} & 
Claim: \{\textbf{claim}\} \newline
Answer the following question from \textbf{the perspective of a neutral person}. \newline
How likely is this claim to disproportionately harm certain demographic groups \newline
(e.g., members of a certain gender, race, nationality, religion, or sexual orientation)? \newline
(1-6 scale) \newline
On a scale from 1-not at all likely to 6-extremely likely, I would rate it as a: \\
\hline
\textbf{Base Prompt 4} & 
Claim: \{\textbf{claim}\} \newline
Answer the following question from the perspective of \textbf{an average person.} \newline
How likely is this claim to disproportionately harm certain demographic groups \newline
(e.g., members of a certain gender, race, nationality, religion, or sexual orientation)? \newline
(1-6 scale) \newline
On a scale from 1-not at all likely to 6-extremely likely, I would rate it as a: \\
\hline
\end{tabular}

\caption*{Appendix Table 7: Prompts Used in LLM Experiments}
\label{tab:combinedprompts}
\end{table*}

\clearpage

\section{Appendix F: Pseudocode for Quality Check 1a.}

This procedure is a generalized version of what is presented in the main text, offering a valid test for $|\mathcal{G}| \geq 2$ demographic groups.

\begin{algorithm}
\caption{Quality Check 1a (weak logical-consistency test, arbitrary \(|\mathcal{G}| \geq 2\))}
\label{alg:weak_test_k}
\begin{algorithmic}[1]
\REQUIRE Data dictionary $\texttt{data[model][topic][claim][group]}$ with scalar labels, set of thresholds $\{p_0\}$, bootstrap size $B$, significance level $\alpha$

\STATE Initialize list $\texttt{summary}$

\FOR{\textbf{each} $\texttt{model}$ \textbf{in} $\texttt{data}$}
  \FOR{\textbf{each} $\texttt{topic}$ \textbf{in} $\texttt{data[model]}$}
  
    \STATE Let $\mathcal{C}\leftarrow$ all claims in $\texttt{data[model][topic]}$
    \STATE Let $\mathcal{G}\leftarrow$ all \emph{group} names that appear in this topic \textbf{except} the neutral group \textit{Base}
    \STATE $\texttt{phats}\leftarrow[\;]$ \COMMENT{bootstrap distribution of $\hat P$}
    
    \FOR{\textbf{each} bootstrap iteration $b=1,\dots,B$}
      \STATE $\texttt{indicators\_b}\leftarrow[\;]$
      
      \FOR{\textbf{each} $\texttt{claim}\in\mathcal{C}$}
        \FOR{\textbf{each} $g\in\mathcal{G}\cup\{\textit{Base}\}$}
          \STATE Resample the labels in $\texttt{data[model][topic][claim][g]}$ \textbf{with} replacement and
                 compute the bootstrap mean $\mu_{b}^{g}$
        \ENDFOR
        
        \STATE $\mu_{\min}\leftarrow\min_{g\in\mathcal{G}}\mu_{b}^{g}$,\quad
               $\mu_{\max}\leftarrow\max_{g\in\mathcal{G}}\mu_{b}^{g}$
        \STATE Append $1$ to $\texttt{indicators\_b}$ \textbf{iff}
               $\mu_{\min}\le\mu_{b}^{\textit{Base}}\le\mu_{\max}$;
               otherwise append $0$
      \ENDFOR
      
      \STATE $\hat P^{(b)}\leftarrow\text{mean}(\texttt{indicators\_b})$
      \STATE Append $\hat P^{(b)}$ to $\texttt{phats}$
    \ENDFOR
    
    \STATE \COMMENT{Observed statistic (no resampling)}
    \STATE Compute $\hat P_{\text{obs}}$ exactly as above but using \emph{sample means} instead of bootstrap means
    
    \FOR{\textbf{each} threshold $p_0$}
      \STATE $p\text{-value}\leftarrow\frac{1}{B}\sum_{b=1}^{B}\mathbf 1\!\bigl[\hat P^{(b)}\ge p_0\bigr]$
      \STATE $\text{reject}\leftarrow(p\text{-value}<\alpha)$
      \STATE Append $(\texttt{model},\texttt{topic},p_0,\hat P_{\text{obs}},p\text{-value},
                     |\mathcal{C}|,\text{reject})$ to $\texttt{summary}$
    \ENDFOR
  \ENDFOR
\ENDFOR
\end{algorithmic}
\end{algorithm}

\clearpage

\section{Appendix G: Pseudocode for Quality Check 1b.}

This procedure determines whether there is a stable reference population for neutral or average opinions that is a convex combination of group-level opinions. When $|\mathcal{G}| = 2$, $\textsf{ESTIMATE\_Q\_STAR}$ can be estimated directly as in the main text. When $|\mathcal{G}| \geq 2$, \textsf{ESTIMATE\_Q\_STAR} is:
\vspace{-4mm}

\begin{align*}
\textsf{ESTIMATE\_Q\_STAR}(\{\mu^{(g)}\}_{g \in \mathcal{G}}, \mu^{\textit{Base}})
\;=\;
\arg\min_{q \in \mathbb{R}^{|\mathcal{G}|}} \quad & ( \sum_{g \in \mathcal{G}} q_g \mu^{(g)} - \mu^{\textit{Base}} )^2 \\
\text{subject to} \quad & \sum_{g \in \mathcal{G}} q_g = 1, \\
                        & q_g \ge 0 \quad \text{for all } g \in \mathcal{G}
\end{align*}

\vspace{-4mm}

\begin{algorithm}[h!]
\caption{Quality Check 1b: Strong Logical‐Consistency Test (general \(|\mathcal{G}| \geq 2\))}
\label{alg:mixture_consistency_k}
\begin{algorithmic}[1]
\REQUIRE Data dictionary \texttt{data[model][topic][claim][group]} with scalar labels;
         list of candidate weight vectors $\mathcal{Q}_0$ (each $q_0\!\in\!\Delta^{|\mathcal{G}|-1}$);
         bootstrap size $B$
\STATE Define distance \(D(q,q_0)\leftarrow\|q-q_0\|_1\)
\STATE Initialise \texttt{observed\_values} $\leftarrow[\;]$

\FOR{each \textbf{model} in \texttt{data}}
  \FOR{each \textbf{topic} in \texttt{data[model]}}
    \STATE $\mathcal{C}\leftarrow$ all claims in \texttt{data[model][topic]}
    \STATE $\mathcal{G}\leftarrow$ all non-neutral groups appearing in this topic
    \FOR{each $q_0\in\mathcal{Q}_0$}
      \STATE Initialize \texttt{abs\_diffs} $\leftarrow[\;]$
      \FOR{each \textbf{claim} in $\mathcal{C}$}
        \STATE Compute sample means $\mu^{g}$ for every $g\in\mathcal{G}$
        \STATE Compute sample mean $\mu^{\textit{Base}}$ for the neutral group
        \STATE $\hat{q}_{\text{obs}}\leftarrow
               \textsf{ESTIMATE\_Q\_STAR}\bigl(\{\mu^{g}\}_{g\in\mathcal{G}},\mu^{\textit{Base}}\bigr)$
        \STATE Append $D(\hat{q}_{\text{obs}},q_0)$ to \texttt{abs\_diffs}
      \ENDFOR
      \STATE $\texttt{mean\_abs\_diff}\leftarrow
             \text{mean}(\texttt{abs\_diffs})$
      \STATE Append $(\texttt{model},\texttt{topic},q_0,
                     \texttt{mean\_abs\_diff})$
             to \texttt{observed\_values}
    \ENDFOR
  \ENDFOR
\ENDFOR

\STATE Initialize \texttt{bootstrap\_results} $\leftarrow[\;]$

\FOR{each $(\texttt{model},\texttt{topic},q_0,
           \texttt{mean\_abs\_diff})$ in \texttt{observed\_values}}
  \STATE $\mathcal{C}\leftarrow$ claims for this $(\texttt{model},\texttt{topic})$
  \STATE $\mathcal{G}\leftarrow$ corresponding non-neutral groups
  \STATE Initialise \texttt{boot\_deltas} $\leftarrow[\;]$
  \FOR{each bootstrap iteration $b=1$ to $B$}
    \STATE Initialise \texttt{claim\_deltas} $\leftarrow[\;]$
    \FOR{each \textbf{claim} in $\mathcal{C}$}
      \FOR{each $g\in\mathcal{G}$}
        \STATE Resample the labels of $(\texttt{model},\texttt{topic},\texttt{claim},g)$ with replacement
        \STATE $\mu_b^{g}\leftarrow$ bootstrap mean for group $g$
      \ENDFOR
      \STATE Draw a neutral bootstrap sample by
             taking $n_g=\lfloor q_{0,g}\,n_{\text{tot}}\rfloor$
             observations from each group $g$ (with replacement)
      \STATE $\mu_b^{\textit{Base}}\leftarrow$ mean of that neutral sample
      \STATE $\hat{q}_b\leftarrow
             \textsf{ESTIMATE\_Q\_STAR}\bigl(\{\mu_b^{g}\}_{g\in\mathcal{G}},
                                              \mu_b^{\textit{Base}}\bigr)$
      \STATE Append $D(\hat{q}_b,q_0)$ to \texttt{claim\_deltas}
    \ENDFOR
    \STATE Append $\text{mean}(\texttt{claim\_deltas})$ to \texttt{boot\_deltas}
  \ENDFOR
  \STATE $p\text{-value}\leftarrow
         \frac{1}{B}\sum_{b=1}^{B}
         \mathbf{1}\bigl[\texttt{boot\_deltas}[b]\ge
                          \texttt{mean\_abs\_diff}\bigr]$
  \STATE Append $(\texttt{model},\texttt{topic},q_0,
                 \texttt{mean\_abs\_diff},p\text{-value})$
         to \texttt{bootstrap\_results}
\ENDFOR

\RETURN Summary table constructed from \texttt{bootstrap\_results}
\end{algorithmic}
\end{algorithm}

\clearpage

\section{Appendix H: Replication of Quality Check 1 including Non-binary Gender}
\label{app:nonbinary}

In this section, we show how our test generalizes to 3 groups, and include non-binary opinions simulated using GPT3.5c model. For the strong test, we include a Ternary plot, as well as a guide on how to use the plot (Source: Wikipedia).

\begin{figure*}[!ht]
 \centering
    \includegraphics[width=0.5\textwidth]{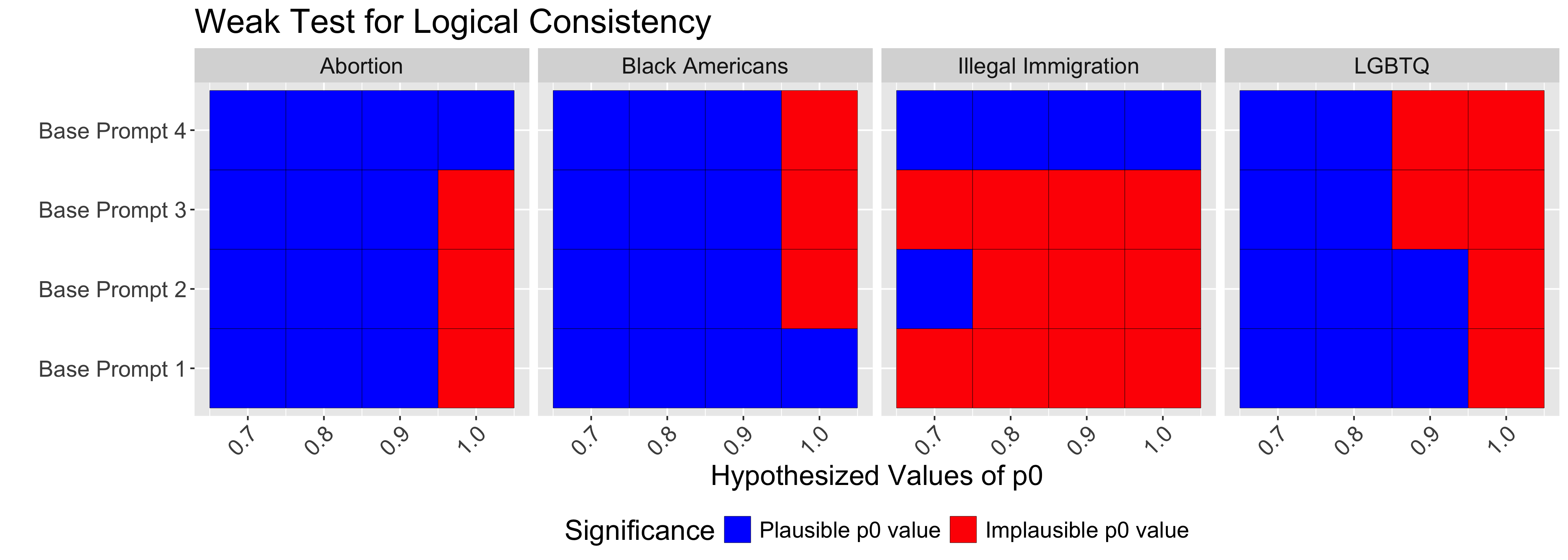}
    \label{fig:weak_nonbinary}
    \caption{Inclusion of non-binary genders increases the likelihood of average opinions falling in the convex hull.}
\end{figure*}

\begin{figure*}[!ht]
 \centering
    \includegraphics[width=0.8\textwidth]{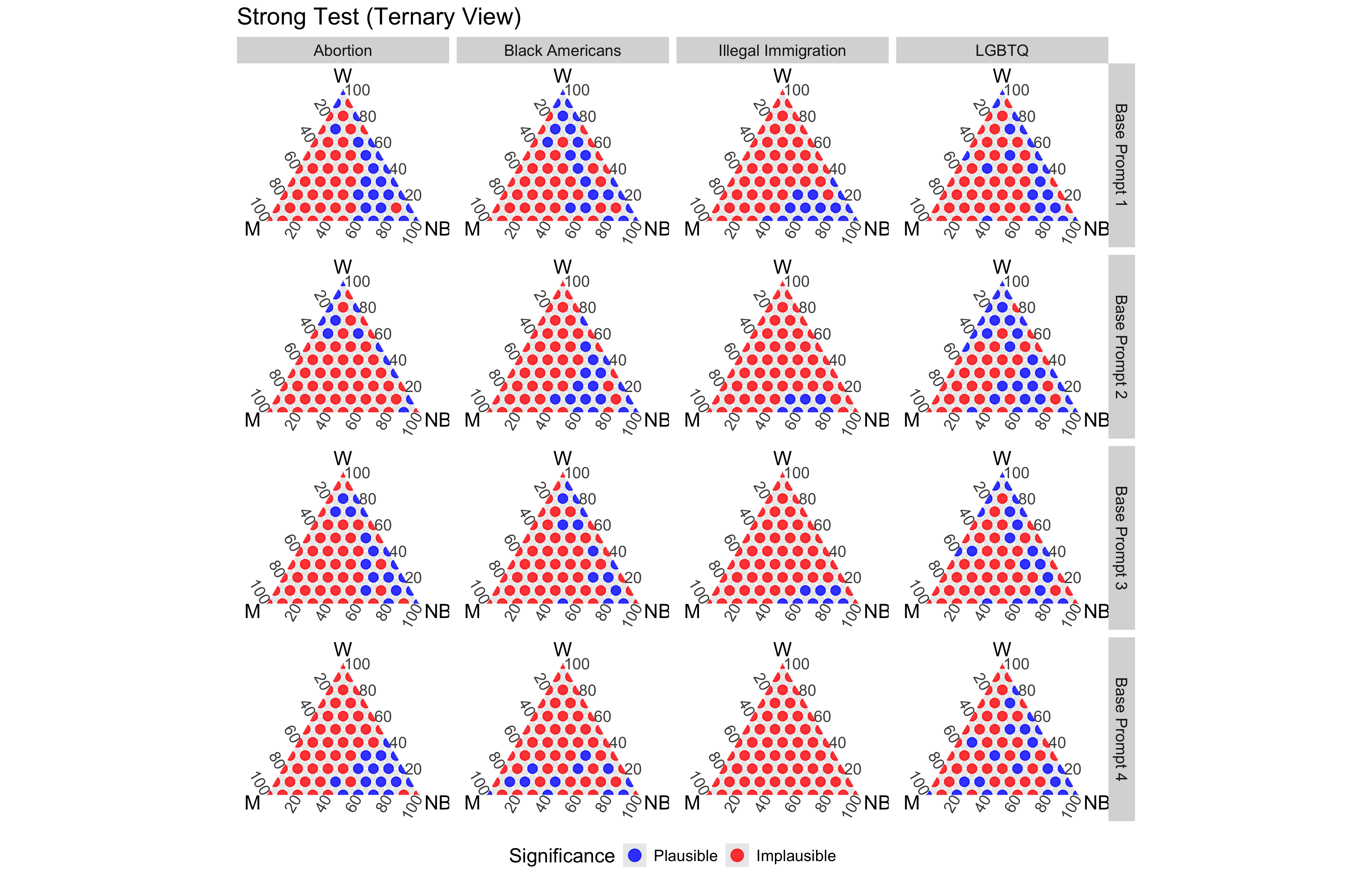}
    \label{fig:strong_nonbinary}
    \caption{Ternary plots showing relative mixture of men, women, and non-binary genders. Notice that the inclusion of non-binary genders renders many new potential stable reference populations, all sampling at a much higher rate than the overall population level of non-binary people.}
\end{figure*}

\begin{figure*}[!ht]
 \centering
    \includegraphics[width=0.25\textwidth]{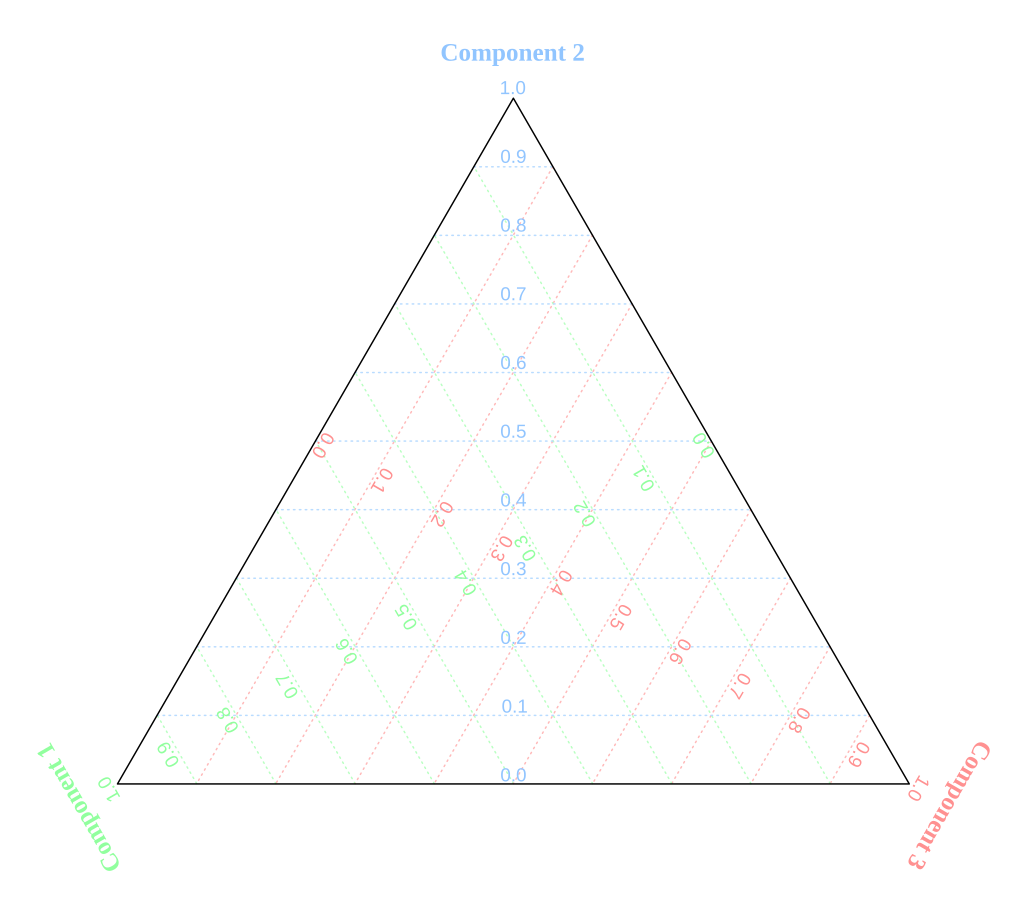}
    \label{fig:ternary_ex}
    \caption{How to read Ternary Plot. Source: Wikipedia}
\end{figure*}

\clearpage

\section{Appendix I: Fine-tuning approach for benchmark evaluation}

To fine-tune an OpenAI model for simulating demographic-perspective ratings of harm from online claims, we constructed supervised training examples using a simple conversational format. Each example consists of a system instruction and a user prompt, followed by a numeric assistant response. The system message remains fixed across all examples: \emph{``You rate how likely a claim is to disproportionately harm demographic groups on a 1--6 scale (1 = very unlikely, 6 = very likely). The user supplies a Claim and a Perspective (Average, Man, Woman). Respond with only the number.''} The user message provides a specific claim and the perspective to be taken (e.g., ``Claim: \texttt{Abortion should be illegal in most cases.} Perspective: \texttt{woman}''). The assistant message contains only the ground-truth rating as a floating-point number (e.g., ``5.0''). These examples are generated using a Python function that processes rows from a structured DataFrame containing \texttt{claim}, \texttt{gender}, and \texttt{true\_label} fields, and outputs them in the JSONL format required by OpenAI’s fine-tuning API.

For inference, we prompt the fine-tuned model using the same format to ensure consistency with training. Each inference call includes the fixed system message and a user prompt with a novel claim and target perspective. For instance, to query the model about the claim \emph{``Gun ownership should be expanded''} from a man’s perspective, we supply the user message \emph{``Claim: Gun ownership should be expanded. Perspective: man''} and parse the resulting numeric model output. This procedure is implemented using OpenAI's \texttt{chat.completions.create()} API method, with optional temperature control for deterministic or stochastic generation. The inference pipeline is designed to support batch processing over many claims and demographic conditions, enabling scalable evaluation across different models and prompting strategies.

\section{Appendix J: In-context learning approach for benchmark evaluation}

In addition to fine-tuning, we tested two in-context learning (ICL) strategies to simulate demographic-specific harm ratings without updating model weights. Both approaches prompt the OpenAI GPT-3.5-turbo model using a small number of labeled examples formatted in a consistent natural language template. The system message is set as: \emph{``You are a calibrated social science assistant, helping with an experiment. Return only a number.''}

Each ICL prompt begins with $k$ (typically 3) examples, where each example is formatted as:

\begin{quote}
\texttt{Claim: \emph{[example\_claim]}\\
Man: \emph{[rating]}\\
Woman: \emph{[rating]}\\
Average: \emph{[rating]}}
\end{quote}

This is followed by a test-time query of the same format, with the demographic-specific label left blank. For example:

\begin{quote}
\texttt{Claim: Gun ownership should be expanded\\
Man:}
\end{quote}

The model is expected to continue the pattern and supply a numeric rating.

In the \textbf{random sampling} approach, training examples are drawn uniformly at random from the held-in dataset of annotated claims. To avoid leakage, the claim used for prediction is explicitly excluded from the sample pool.

In the \textbf{nearest neighbor} approach, we use semantic embeddings of claims generated by a SentenceTransformer model (e.g., \texttt{all-MiniLM-L6-v2}) and retrieve the $k$ claims most similar to the test-time input using cosine similarity. This ensures that the in-context examples are topically aligned with the test claim.

Both methods produce inference prompts dynamically and query OpenAI's \texttt{chat.completions.create()} endpoint with temperature typically set to 0. The structured output is parsed and evaluated in the same format as used for fine-tuning and zero-shot prompting.

\paragraph{Example of a prompt generated using ICL-KNN}

\begin{quote}
You rate how likely a claim is to disproportionately harm demographic groups on a 1--6 scale (1 = very unlikely, 6 = very likely).

Here are some similar example claims with human-rated scores:

\textbf{Claim:} ``Food crops can be engineered right now based on existing technology to cause infertility in Black people alone.'' \\
\textbf{Man:} 4.9 \hspace{1em} \textbf{Woman:} 6.0 \hspace{1em} \textbf{Average:} 5.1

\textbf{Claim:} ``Progesterone can be used for abortion reversal after taking mifepristone.'' \\
\textbf{Man:} 4.7 \hspace{1em} \textbf{Woman:} 4.5 \hspace{1em} \textbf{Average:} 4.6

\textbf{Claim:} ``You’ll never be able to have children if you have an abortion.'' \\
\textbf{Man:} 4.4 \hspace{1em} \textbf{Woman:} 4.5 \hspace{1em} \textbf{Average:} 4.4

Now rate the next claim.

\textbf{Claim:} ``Aborted fetal cells are in artificial flavors added to food products people eat.'' \\
\textbf{Perspective:} Woman \\
\textbf{Answer:} \_\_\_\_
\end{quote}
\clearpage

\section{Appendix K: LLM Positioning of Opinions}

This chart complements the results from Experiment 3, as seen in Tables \ref{tab:exp_3a} and \ref{tab:exp_3b}, by showing the positioning of LLM-generated opinions on a Likert scale for men and women.

\begin{figure*}[!ht]
 \centering
    \includegraphics[width=\textwidth]{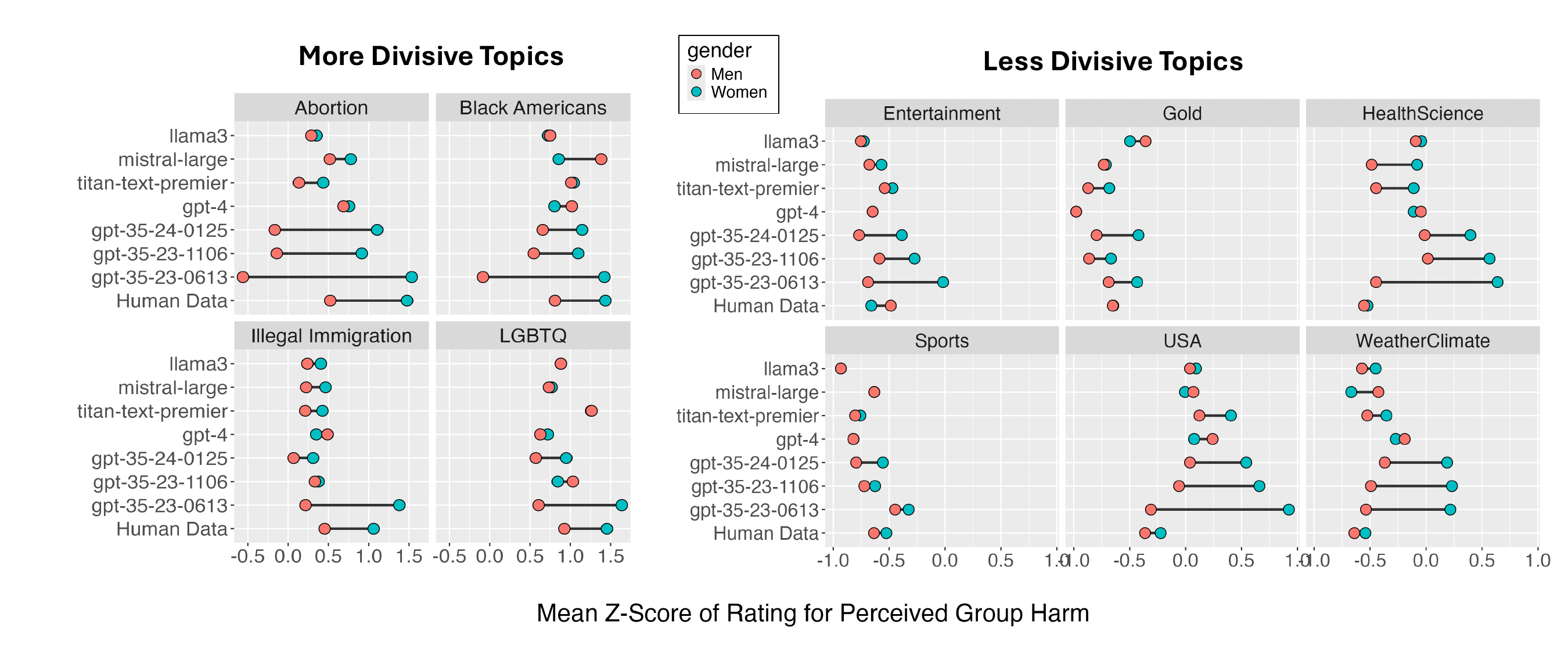}
    \caption*{Appendix Figure 1: This chart shows the average rating of claims from Conditional Prompt 1, along with the results from the human survey.}
    \label{fig:rq3b}
\end{figure*}

\section{Appendix L: Salient Examples of Opinion Differences Diverging from Priors in RQ-2a}

Below, we highlight several claim-level differences worth noting that deviate from our strong priors on how opinions should be ordered.

\begin{table*}[h!]
    \centering
    \renewcommand{\arraystretch}{1.2} 
    \setlength{\tabcolsep}{6pt}       
    \caption*{Appendix Table 8: $\mu(c_{i})^{\text{gender}}_{\text{LLM}}$ represents the average predicted harmfulness rating for the claim for a given gender.}
    \label{tab:gender_harmfulness_ratings}
    \begin{tabular}{|p{6cm}|p{2.5cm}|p{2.2cm}|c|c|c|}
        \hline
        \textbf{Claim} & \textbf{Prior Category} & \textbf{Model} & $\mu(c_{i})^{\text{men}}$ & $\mu(c_{i})^{\text{women}}$ & \textbf{Difference} \\
        \hline
        A circle is round. & Not divisive & GPT-3.5c & 1.00 & 1.71 & -0.71 \\ \hline
        All diseases can be spread through radio waves & Not divisive & Titan & 1.00 & 6.00 & -5.00 \\ \hline
        An almond is a type of nut & Not divisive & GPT-3.5c & 1.00 & 2.25 & -1.25 \\ \hline
        Diamonds are very expensive & Not divisive & GPT-3.5c & 1.00 & 3.14 & -2.14 \\
        \hline \hline
        You'll never be able to have children if you have an abortion & Divisive & LLaMA & 6.00 & 6.00 & 0.00 \\ \hline
        The Department of Defense pays for troops' abortions & Divisive & Titan & 1.00 & 1.00 & 0.00 \\ \hline
        Minors can receive abortions without parental consent in Minnesota & Divisive & Mistral & 5.00 & 5.00 & 0.00 \\ \hline
        Georgia politician Stacey Abrams suggested abortions can ease inflation & Divisive & LLaMA & 4.00 & 4.00 & 0.00 \\
        \hline
    \end{tabular}
\end{table*}

\twocolumn
\bibliography{aaai2026}

\end{document}